\documentclass[%
reprint,
 amsmath,amssymb,
 aps,
pra,
]{revtex4-2}
\usepackage[T1]{fontenc}
\usepackage{graphicx}
\usepackage{dcolumn}
\usepackage{bm}

\begin{document}

\preprint{APS/123-QED}

\title{ Gauge Symmetry as the Origin of Spin–Orbit Coupling
of Light}

\author{Lili Yang$^{1}$}
\author{Longlong Feng$^{1}$}
\email{flonglong@mail.sysu.edu.cn}
\author{Pengming Zhang$^{1}$}
\email{zhangpm5@mail.sysu.edu.cn} 
\affiliation{$^{1}$ School of Physics and Astronomy, Sun Yat-Sen University, Zhuhai 519082, China}

\date{\today}

\begin{abstract}
Dual symmetry is an intrinsic property of Maxwell's equations, corresponding to a global U(1) symmetry in vacuum, with helicity as the associated conserved quantity. In this paper, we investigate light propagation in a helical fiber system using a field-theoretical approach and introduce an effective gauge field  that emerges from the localization of dual symmetry. We show that the helical trajectory of light rays reveals this gauge field as a manifestation of spin–orbit coupling. Although orbit–orbit coupling also arises in such systems, the spin–orbit interaction possesses deeper physical significance, as it originates from the intrinsic dual symmetry embedded in Maxwell's equations.
\end{abstract}
    \maketitle
\section{Introduction}
Symmetry is a cornerstone of modern physics. According to Noether’s theorem, every continuous symmetry in a system corresponds to a conserved quantity. For the electromagnetic field, a particularly important symmetry is duality symmetry—the invariance of Maxwell’s equations in vacuum under the exchange of electric and magnetic fields ~\cite{Jackson1999ClassicalE3}. This dual symmetry becomes especially meaningful in the presence of magnetic charges, where it can be formulated in a fully covariant, relativistic framework using four-vectors ~\cite{Zwanziger1968QUANTUMFT, Han1971ManifestDI, Gambini1979NullplaneAO, Singleton1997MagneticCA, Kato2001GaugingDS, Arrays2017SpinOrbitalMD}. Within this context emerges the concept of “electromagnetic democracy” ~\cite{berry2002optical, Bliokh2013}—a viewpoint in which electric and magnetic fields are treated on equal footing. In vacuum, this symmetry is embodied by the discrete transformation $({\bf E}, {\bf B}) \rightarrow ({\bf B}, -{\bf E})$, under which Maxwell’s equations remain unchanged. More generally, this duality can be expressed as a $U(1)$ rotation in the real $\left({\bf E}, {\bf B}\right)$ plane:
\begin{equation}\label{eq:EB-duality}
\left(\begin{array}{c} {\bf E} \\{\bf B} \end{array}\right) \rightarrow\left(\begin{array}{cc}
\cos \theta & \sin \theta \\
-\sin \theta & \cos \theta
\end{array}\right) \left(\begin{array}{c} {\bf E} \\{\bf B} \end{array}\right)=R(\theta)\left(\begin{array}{c} {\bf E} \\{\bf B} \end{array}\right),
\end{equation}
where $\theta$ is the duality rotation angle. The classical "electromagnetic democracy" is given at $\theta=\pi/2$. Following Eq.(\ref{eq:EB-duality}), the electromagnetic field exhibits a global $U(1)$ symmetry, ensuring that helicity is conserved, as dictated by Noether’s theorem ~\cite{Calkin1965}. Of course, a natural question arises: what are the implications if the aforementioned dual transformation is localized? We will show that in helical fibers, the dual symmetry can indeed be localized in an optical-path-dependent manner, thereby ensuring the conservation of helicity ~\cite{FernandezCorbaton2013}.
We promote the global dual transformation in Eq.~(\ref{eq:EB-duality}) to a local one in terms of a phase transformation of complex fields:
\begin{equation}\label{DU}
{\bf D}\pm i{\bf B} \rightarrow \text{e}^{\mp i\theta({\bf x})} ({\bf D}\pm i{\bf B}),
\end{equation}
where $\theta({\bf x})$ is a function of coordinate ${\bf x}$. In the presence of a medium, the dual transformation should replace $\mathbf{E}$ with the electric displacement field $\mathbf{D}$ to maintain the source-free condition. Moreover, the corresponding rotation angle now depends on spatial coordinates, meaning that the original global $U(1)$ symmetry may become a $local\, U(1)$ symmetry.

We carry out our analysis in helicity space, defined by a transverse basis relative to a chosen reference direction—typically aligned with the incident light’s propagation direction. Without loss of generality, we take this to be the $z$-axis. The helicity eigenvectors satisfy: $({\bf e}_{z}\cdot{\bf s}){\bf e}_{\pm}=\pm {\bf e}_{\pm}$ where ${\bf s}=\{s_i\}^{jk}=-i\epsilon_{ijk}$ is the adjoint representation of spin-1 matrices. In this helicity basis, the transverse components of the electric displacement and magnetic field are expressed as $\mathcal{D}_{\perp}=(D_{+},D_{-})^{T}$ and $\mathcal{B}_{\perp}=(B_{+},B_{-})^T$. Furthermore, we introduce the 2-vector  wavefunctions of the photon in the transverse plane as ~\cite{Feng2022FourvectorOD}
\begin{equation} \label{WF} 
\psi^{\pm}= \mathcal{D}_{\perp} \pm i \sigma_3 \mathcal{B}_{\perp}. 
\end{equation}

In vacuum, $\psi^{\pm} $ represent the components of positive and negative energy states in the helicity basis, respectively~\cite{Keller2005}. A negative-energy solution can be interpreted as an antiparticle traveling backward in time along its worldlines. However, since the photon is its own antiparticle, an antiphoton can be interpreted as a photon mode propagating in the opposite direction. In our previous analysis, we found that in general media, the helical state undergoes nontrivial transformations between these two modes and evolves independently as its own degree of freedom~\cite{yang2025_inducedBerry}.

By introducing a four-component wave function, $\psi =  (\psi^+ \ \psi^-)^{\mathrm{T}}$, Maxwell's equations can be recast into an optical Dirac-like form when interactions with a medium are taken into account ~\cite{Feng2022FourvectorOD}. Under the local dual transformation (\ref{DU}), the four-component wave function transforms as:
\begin{equation}
    \begin{aligned}\label{eq:spin_dual}
        \psi \rightarrow e^{-i\gamma^0\hat{S}_z\theta(\mathbf{x})}\psi,
    \end{aligned}
\end{equation}
where the helicity operator $\hat{S}_z$ is defined as
\begin{equation}
    \begin{aligned}\label{Spin}
        \hat{S}_z =\left(\begin{array}{cc}
            \sigma_3 &   \\
             & \sigma_3 
        \end{array}\right) .
    \end{aligned}
\end{equation}

\section{Dual gauge field}
According to the principles of gauge theory ~\cite{YangMills1954}, preserving a local symmetry requires the introduction of an associated gauge field. From an optical perspective, the transformation in Eq.~(\ref{eq:spin_dual}) may be viewed as a position-dependent rotation in the internal (spin/helicity) space of the field. When the transformation parameter $\theta(\mathbf{x})$ varies in space, taking a spatial derivative necessarily acts on both the field and the parameter, thereby producing an extra term that cannot be removed by the ordinary derivative alone. For an infinitesimal transformation, this reads
\begin{equation}
 \delta (\partial_i \psi) = -i\gamma^0\hat{S}_z \partial_i(\theta(\mathbf{x})\psi ).
\end{equation}
Equivalently, the derivative generates an additional contribution proportional to $\partial_i\theta(\mathbf{x})$, i.e., an effective “connection” acting in the helicity (spin) subspace. To preserve gauge invariance and compensate for this extra term, we introduce a \textbf{dual gauge field} $A_i$ and promote the ordinary derivative to a covariant derivative,
\begin{equation}\label{covariant derivative}
    \partial_i \rightarrow D_i, \quad D_i = \partial_i + i\gamma^0\hat{S}_z \mathcal{A}_i, \; \rightarrow \; \hat{K}_i = \hat{k}_i + \gamma^0\hat{S}_z \mathcal{A}_i,
\end{equation}
where $\hat{K}_i$ denotes the momentum-operator representation of the covariant derivative. The gauge field $A_i$ is determined by, and encodes, variations of the medium. In Sec. \ref{SOI}, we derive an explicit representation of the coupling factor $\gamma^0\hat{S}_z$in terms of the photon helical eigenfunctions.in terms of the photon helical eigenfunctions. In the present context, this coupling constant is associated with the photon helicity and thus corresponds to a conserved quantity protected by dual symmetry, in contrast to the electric charge $e$ that sets the coupling strength in electrodynamics.

The total Lagrangian can thus be written as
\begin{equation}\label{Total lagrangian}
\mathcal{L} = \mathcal{L}_0 + \mathcal{L}_{\mathcal{A}},
\end{equation}
where $\mathcal{L}_0$ represents the free photon part associated with global dual symmetry and can be constructed from the corresponding optical Dirac equation ~\cite{Feng2022FourvectorOD}, while $\mathcal{L}_{\mathcal{A}}$ accounts for the interaction between the photon and the gauge field, arising from the localization of the symmetry via $\theta(x)$.

For a spin-degenerate symmetric medium, excluding the spatial derivative terms of constitutive tensors yields the zeroth order Lagrangian,
\begin{equation}\label{eq:Lagrangian_zeroth}
    \mathcal{L}_0 = \overline{\psi}(i\gamma^{0}\partial_{0} - \hat{m}_0  
  + \gamma^0 \boldsymbol{\gamma} \cdot \hat{\bf p}^0)\psi,
\end{equation}
where $\overline{\psi} =\psi^{\dagger}\gamma^0$ is the conjugated field, $\gamma^{\mu}$ are the gamma matrices in the Dirac representation and $\hat{\mathbf{k}} = -i\bm{\nabla}$ is the momentum operator. $\hat{m}_0$ and $\hat{\bf p}^0=\{{\bf p}_{\perp}^0,p^0_z\}$ denote the effective mass and momentum operators, 
\begin{eqnarray}
        &\hat{m}_0 =  Q_0\hat{k}_z -\Delta \hat{m}_0, \quad \Delta\hat{m}_0=\mathbf{q}\cdot  \hat{\mathbf{k} }_{\perp} - \displaystyle{\frac{q_0}{\hat{k}_z}} \hat{k}_+  \hat{k}_{-} , \label{eq:m0} \\ 
        &\hat{p}^0_{ \pm}= Q_{ \pm 2}\hat{k}_z- 2Q_{ \pm 1}\hat{k}_{ \pm} +\displaystyle{\frac{q_0}{k_z}} \hat{k}_{\pm}^2, \quad p_z^0=0. \label{eq:p0} 
\end{eqnarray}
In the helicity space, the transverse components are $\hat{\mathbf{k}}_{\perp} = (\hat{k}_+, \hat{k}_-)^T$ with $\hat{k}_{\pm}=(\hat{k}_x\mp i\hat{k}_y)/\sqrt{2}$; the multipole moments are related to inverse permittivity/permeability tensors $\epsilon^{-1}=\mu^{-1}=\{\Pi_{ij}\}$ through a unitary transformation $\hat{U}$,
\begin{equation}
    \begin{aligned}\label{jiezhizhangliang}
\hat{U}^{-1} \Pi \hat{U}=\left(\begin{array}{ccc}
Q_0 & Q_{+2} & Q_{+1} \\
Q_{-2} & Q_0 & Q_{-1} \\
Q_{-1} & Q_{+1} & q_0
\end{array}\right),
    \end{aligned}
\end{equation}
where the decomposition leads to two monopole moments $Q_0=\frac{1}{2}(\Pi_{11}+{\Pi}_{22})$ and $q_0=\Pi_{33}$, dipole $Q_{\pm 1}=\frac{1}{\sqrt{2}}(\Pi_{13}\mp i\Pi_{23})$ that form a dipole vector ${\bf q}=\{Q_{+1},Q_{-1}\}$, and quadrupole $Q_{ \pm 2}=\frac{1}{2}\left[\left(\Pi_{11}-\Pi_{22}\right) \mp i\Pi_{12}\right]$, corresponding to the scalar (spin-0), vector (spin-1), and tensor (spin-2) modes, respectively~\cite{Feng2022FourvectorOD}.

\subsection{Pure Gauge Potential}
By replacing the original momentum operator $\hat{k}_i$ in the Lagrangian $\mathcal{L}_0$ with $\hat{K}_i$, we obtain the total Lagrangian $\mathcal{L}$ that includes the contribution from the gauge field,
\begin{equation}\label{eq:Lagrangian_total}
    \mathcal{L} = \overline{\psi}(i\gamma^{0}\partial_{0} - \hat{M}_0  + \gamma^0 \boldsymbol{\gamma} \cdot \hat{\bf P}^0)\psi,
\end{equation}
where 
\begin{eqnarray}
        &\hat{M}_0 =  Q_0\hat{K}_z -\Delta \hat{M}_0, \quad \Delta\hat{M}_0=\mathbf{q}\cdot  \hat{\mathbf{K} }_{\perp} - \displaystyle{\frac{q_0}{\hat{K}_z}} \hat{K}_+  \hat{K}_{-} , \label{eq:M0} \\ 
        &\hat{P}^0_{ \pm}= Q_{ \pm 2}\hat{K}_z- 2Q_{ \pm 1}\hat{K}_{ \pm} +\displaystyle{\frac{q_0}{K_z}} \hat{K}_{\pm}^2, \quad P_z^0=0. \label{eq:P0} 
\end{eqnarray}
It can be verified that the system remains invariant under the gauge transformations given in Eq.(\ref{eq:spin_dual}). When a local transformation parameter  $\theta (\bm x)$ is introduced, substituting this transformation into the Lagrangian in Eq.(\ref{eq:Lagrangian_total}) shifts all momentum operators according to
\begin{equation}
    \hat{k}_i \rightarrow \hat{k}_i - \gamma^0S_z \partial_i\theta.
\end{equation}
Accordingly, the gauge field  transform as $\mathcal{A}_i\rightarrow \mathcal{A}_i +\partial_i \theta$ so as to compensate for the coordinate-dependent variation.

Retaining only the leading-order contribution, the interaction between the gauge field and the photon field can be written as
\begin{equation}\label{L_A}
    \mathcal{L}_A = \overline{\psi}\{-\gamma^0\hat{S}_z(Q_0\mathcal{A}_z  + \bm q\cdot \mathbf{\mathcal{A}}) + \gamma^{\pm} \hat{S_z} (Q_{\pm2} \mathcal{A}_z -2Q_{\pm1} \mathcal{A}_{\pm}  )\}\psi.
\end{equation}
Here, the effects of spatial variations in the material tensor are encoded in the \textbf{dual gauge field} $\mathcal{A}_i$.
This expression describes an additional light–matter interaction mediated by the elementary excitations in terms of multipole modes, the induced gauge field, and the light field.

Alternatively, one can consider photon dynamics in an optical medium without introducing a gauge field. Thus, the Lagrangian becomes 
\begin{equation}
    \mathcal{L} = \mathcal{L}_0 + \mathcal{L}_Q
\end{equation}
where $\mathcal{L}_Q$ captures terms involving first-order spatial derivatives~\cite{yang2025_inducedBerry},
\begin{equation}\label{L_Q}
     \mathcal{L}_Q = \overline{\psi}( - \hat{m}_q  
  + \gamma^0 \boldsymbol{\gamma}\cdot \hat{\bf p}^q )\psi,
\end{equation}
with
\begin{eqnarray}
   \hat{m}_q &=& \hat{k}_zQ_0+\frac{i}{2}{\bm \nabla}_{\perp} \cdot \mathbf{q} - \hat{L}^{z}_q\hat{S}_z,\label{mI} \\
    \hat{p}^{q}_{\pm} &=& \hat{k}_z Q_{ \pm 2}- 2 \hat{k}_{ \pm}Q_{ \pm 1} +\frac{\hat{k}_{\pm}q_0}{k_z} \hat{k}_{ \pm},\quad \hat{p}^q_z=0 .\label{pI}
\end{eqnarray}
The operator $\hat{L}^z_0$ exhibits the characteristic features of orbital angular momentum and is defined as
 \begin{equation}
    \hat{L}_q^z= \hat{\bf L}_q\cdot{\bf e}_z=\frac{1}{2}({\bm \nabla}_{\perp}\times \mathbf{q})\cdot \bm{e}_z.
\end{equation}
More generally, to account for the curved trajectory of light, we reformulate the Maxwell equation in a curvilinear coordinate with metric $g = \{g_{ij}\}$. Given the fact that curved space is equivalent to a spin-degenerate medium with the constitutive tensors
$\epsilon_g$ and $\mu_g$ obeying the simple equality to the spatial metric, i.e., $\epsilon_g^{-1} = \mu_g^{-1} = g$, we can have effective constitutive tensors simply by matrix multiplication, $\epsilon\rightarrow \epsilon_g\epsilon$, $\mu\rightarrow \mu_g\mu$ ~\cite{WuFeng_GW2022, yang2025_inducedBerry}. 

When both the curl and the divergence of the dipole moment in Eq.~(\ref{mI}) vanish, the corresponding local gauge field becomes trivial. In this case, the gauge potential takes the pure-gauge form $\mathcal{A}_i=iU^{\dagger}\nabla_iU$, where the transformation factor take $U=e^{-i\gamma^0\hat{S}_z\theta(\mathbf{x})}$. This corresponds to a vanishing gauge field strength, $\mathcal{F}_{ij}=0$. By identifying Eqs.~(\ref{L_A}) and (\ref{L_Q}), we find that the following condition must hold:
\begin{equation}
\hat{k}_i Q_j = \gamma^0\hat{S}_zQ_j \mathcal{A}_i,
\end{equation}
where $Q_j$ collects all components of the medium tensor appearing in Eq.~(\ref{jiezhizhangliang}). Solving this relation yields
\begin{equation}
Q_j = Q_j^c \, e^{i\gamma^0 \hat{S}_z \int \mathcal{A}_i \, d x^i},
\end{equation}
where $Q_j^c$ denotes a constant amplitude. At the local scale, the perturbations induced in the medium by incident light depend on its polarization. Distinct polarization states generate dipole responses with different orientations, implying that the polarization of light governs the dipole-induced modification of the medium, expressed as
\begin{equation}
    \triangle Q_j \approx \lambda \mathcal{A}_i  \triangle x^i, 
\end{equation}
where $\lambda$ represents the polarization (helicity) index of the light.

\subsection{Nontrivial Dual Gauge Field}
Since the helical eigenstate of light serves as an independent propagation quantity under spin-degenerate conditions, we can formulate the $U(1)$ dual symmetry by taking the helical eigenstate as the fundamental element.
Let $\psi_i$ denote the $i$th-component of a four-vector field $\psi$. The constraint imposed by helicity conservation allows the wavefunction to be decomposed into two opposite helicity states $\psi_\sigma$ with the helicities $\lambda=\pm 1$: $\psi_{+1} = (\psi_1,0,0,\psi_4)^T$, and $\psi_{-1}=(0,\psi_2, \psi_3,0)^T$.  
Under a local dual transformation, the helicity component $\psi_\lambda$ transforms as 
\begin{equation} \label{eq:helicity-dual}
     \psi_\lambda \rightarrow e^{i\lambda\theta(\mathbf{x})}\psi_\lambda,\quad \lambda=\pm 1.
\end{equation}
Accordingly, the transformation in Eq.~(\ref{eq:spin_dual}) reflects a helicity rotation symmetry, which manifests as a $U(1)$ subgroup of the spin group $SU(2)$, in the form $e^{i\boldsymbol{\sigma} \cdot \mathbf{n}}$. As the topological charge $l$ does not affect the fundamental dual symmetry, we will set aside its discussion for now and revisit the case of vortex light at a later stage. 

In this helicity representation, the coupling term $\gamma^0\hat{S}_z$ can be expressed directly in terms of the helicity eigenvalue $\lambda$.  Furthermore, one can verify that, when Eq.~(\ref{eq:Lagrangian_total}) is transformed to this basis, the total Lagrangian becomes block-diagonal, decomposing into two independent $2\times 2$ diagonal blocks. This structure implies that the helicity eigenstates decouple and their corresponding wavefunctions evolve independently.

The effective gauge field experienced by photons in the medium is thus directly determined by the spatial derivatives of the medium. Instead of constructing the Lagrangian term by term to cancel variations~\cite{Kato2001GaugingDS}, we focus on the regime where diffraction is negligible, so that the localization of the duality symmetry depends only on the optical path length $s$, i.e., through the phase angle $\theta(s)$ in the dual transformation, see \eqref{eq:spin_dual} and \eqref{eq:helicity-dual}.

For photonic systems, the arc length parameter $s$ can be regarded as a time parameter in natural units, and the Lagrangian takes the form
\begin{equation}\label{bianfen}
    \mathcal{L} = \overline{\psi }(i\gamma^{0}D_{0} -\hat{m}_0  
  + \gamma^0 \boldsymbol{\gamma} \cdot \hat{\bf p}^0) \psi,
\end{equation}
The covariant derivative $D_0$ with respect to time can equivalently be expressed in terms of the path length derivative $D_s$.

In the helical representation $\psi_{\lambda}$, the only nonvanishing component of the gauge field appears along the propagation path, yielding
\begin{equation}\label{covariant}
    D_s = \partial_s + i \lambda {\mathcal{A}}_s.
\end{equation}
Under the local transformation introduced in Eq.~(\ref{eq:spin_dual}), the gauge field transforms as
\begin{equation}
   {\mathcal{A}}_s \rightarrow {\mathcal{A}}_s + \partial_s\theta(s) ,
\end{equation}
leaving the Lagrangian invariant.
The covariant-derivative structure can be naturally extended to momentum space by introducing
\begin{equation}
    \begin{aligned}
        D_s = \partial_s + i \lambda {\mathcal{A}}_s \rightarrow  \frac{d k^i}{ds}D^k_{i} =\frac{d k^i}{ds }(\frac{\partial}{\partial k_i} + i \lambda {\mathcal{A}}_i(k) ) ,
    \end{aligned}
\end{equation} 
or equivalently,
\begin{equation}
    D_i^k = \partial^k_i + i \lambda {\mathcal{A}}_i(k).
\end{equation}
We introduce the following momentum-space representation by making the substitution:
\begin{equation}\label{momentum-space representation}
      R_i\rightarrow \partial^k_i,\quad r_i \rightarrow D^k_i,
\end{equation}
where $R$ denotes the canonical coordinate and $r$  the physical coordinate. They are related through the gauge field  $\bm A(k)$ via $\bm r=\bm R - \bm A(k)$~\cite{Bliokh2004SpinGF,yang2025_inducedBerry}. 
The Lagrangian associated with the corresponding covariant derivative in Eq.~(\ref{bianfen}) is 
\begin{equation}\label{spin-orbit}
    \mathcal{L} = -\sum_{\sigma} \psi_{\lambda}^{\dagger} \dot{\mathbf{k}}\cdot D^k_i \psi_{\lambda}.
\end{equation}
Expanding the wavefunction in momentum modes and replacing $\psi_{\lambda}(r)$ by $\psi_{\lambda}(k)$, one obtains—in the semiclassical approximation—the momentum-space Lagrangian for particle motion, 
\begin{equation}
    \mathcal{L}_{semi}=\dot{\bm k} \cdot \bm r(k).
\end{equation}

Applying the Euler–Lagrange equations to this Lagrangian yields the semiclassical equation of motion
\begin{equation}
    \dot{\bm r} \sim \lambda \dot{\bm k}\times \bm F,
\end{equation}
with $\bm F=\bm {\nabla_k} \times \bm A$. This relation provides the semiclassical description of the spin Hall effect of light~\cite{Bliokh2008GeometrodynamicsOS,Bliokh2009}. The \textbf{dual gauge field}, commonly identified with the Berry connection,  has an associated Berry curvature that in momentum space takes the form
\begin{equation}
    \bm F = \lambda\mathbf{k}/k^3,
\end{equation}
i.e., a “magnetic monopole” configuration~\cite{Chiao1986ManifestationsOB,BialynickiBirula1987BerrysPI,Bliokh2006}. These ideas have been applied to explain a variety of optical phenomena, including Rytov’s law ~\cite{KravtsovOrlov2019,Chiao1986ManifestationsOB}, as well as spin-dependent transverse shifts such as the Imbert--Fedorov shift and the spin Hall effect of light ~\cite{Fedorov2013ToTT,Imbert1972CalculationAE,Murakami2003DissipationlessQS,Onoda2004HallEO,Mathieu2004,Sinova2004,Bliokh2004SpinGF,Bliokh2005ConservationOA,Duval_2006,Bliokh2006,Bliokh2015SpinorbitIO}.

In the framework of noncommutative mechanics~\cite{Jackiw2000AnyonSA,Duval2001ExoticGS,Duval_2005}, the coordinate $r_i$ 
ceases to be commutative and instead obeys 
\begin{equation}
    [r_i,r_j] = i\lambda \mathcal{F}_{ij}.
\end{equation}
This noncommutative structure is closely connected to the photon-coordinate algebra introduced by Schwinger in his analysis of the photon-localization problem~\cite{Schwinger1998Particles,2025JMP....66h2302S}. It therefore suggests that the issue of photon localization is fundamentally associated with the intrinsic \textbf{dual gauge field}. 

By close analogy with an electron in an external electromagnetic field, the canonical momentum  $\bm P$ and the mechanical momentum $\bm p$ are related through the vector potential as $\bm P - e \widetilde{\bm A}(\bm r) = \bm p$. The associated momentum operators therefore obey the nonvanishing commutation relation
\begin{equation}
    [p_i,p_j] = ie \epsilon_{ijk} B^k,
\end{equation}
with $\bm B = \bm \nabla \times \widetilde{\bm A}$. In the presence of a magnetic monopole of charge $g_m$, for which $\bm B = g_m \bm r / r^3$, a dual correspondence arises between the helicity $\lambda$ and the electric and magnetic charges~\cite{2025JMP....66h2302S}.

The Berry phase can also be understood as a gauge-invariant Wilson loop in momentum space, arising from the integral of the field strength along a closed loop in momentum space~\cite{Milione2012HigherOP,Bliokh2006}. The Wilson loop is given by
\begin{equation}
    U_g =\exp[i\lambda\oint_{\Gamma} dk^i A_i(k) ] = \exp[i\lambda\oint_{\Sigma} dS^{ij} \mathcal{F}_{ij}(k) ],
\end{equation}
where $\Gamma$ is a closed path in momentum space, the $\Sigma$ is the surface that spans the closed loop $\Gamma$, and the $dS^{ij}$ represents an area element.

\section{Gauge field representation of spin-orbit coupling}\label{SOI}
In this section, we will examine how the spatial rotation of the dipole term $\bm q$ along the helical trajectories of photons induces nontrivial gauge fields, and show that these dipole-driven effects constitute the physical origin of the gauge fields associated with spin–orbit coupling.

We focus only on the Lagrangian of the \textbf{dual gauge field} in coordinate space. In the helicity-eigenstate representation $\psi_{\lambda}$, the effective Lagrangian associated with the gauge-field component $A_s$ is 
\begin{equation}\label{eq:Lagrangian_A}
    \mathcal{L}_A = -\sum_{\lambda} \lambda\psi^{\dagger}_{\lambda}  {\mathcal{A}}_s \psi_{\lambda},
\end{equation}
which differs from the standard quantum electrodynamics (QED) interaction. Within our optical–Dirac framework, we formulate a QED-like theory for photons, in which the photon is described by a Dirac-like spinor field $\psi_{\lambda}$, and the $U(1)$ gauge field is induced by the material system—specifically, the spatial derivatives of the dipole components $\bm q$ —with the coupling strength determined by the helicity $\lambda$ of the light. 

To describe the local geometry of the optical path, the Frenet coordinate system based on a moving orthonormal frame is used  ~\cite{Takagi1992QuantumMO,Alexeyev2006TopologicalPE,yang2025_inducedBerry}. In this setting, the induced material tensor parameterized by the multipole moments reduces to the constant components, $Q_0 =1$, $Q_{\pm2}=0$, and the spatially varying components, $Q_{\pm 1}=\mp i \tau r_{\pm}$, $q_0 = 1+\tau^2 r^2$, where $r_{\pm} = (x \mp i y)/\sqrt{2}$, $\tau$ is the torsion of the optical path. The dipole field $\mathbf{q}$ corresponds to a rotation vector in real space, ${\bf q}=\boldsymbol{\Omega}_{\tau}\times{\bf x}$ with $\boldsymbol{\Omega}_{\tau}=\tau {\bf e}_z$. Here, $\tau$ is a slowly varying spatial parameter and may be treated adiabatically as a constant, leading to $\hat{k}_+ Q_{+1} = \hat{k}_- Q_{-1} = 0$. Similarly, $\hat{k}_{\pm}q_0\approx 0$, and $\hat{\mathbf{k}}_{\perp} \cdot \mathbf{q}  = 0$, implying ${\bf q}$ is a divergence-free vector. 

In addition to the longitudinal propagation term $\hat{k}_z$ in the effective mass, two correction terms arise from light–medium interaction: $\Delta\hat{m}_0$, associated with the homogeneous medium response, and $\hat{m}_q$, originating from the non-commutative nature of the momentum operator with spatially varying constitutive tensors. Neglecting the paraxial term, the first correction takes the form $\Delta\hat{m}_0 = \mathbf{q}\cdot \hat{\mathbf{k}}_{\perp} = \hat{L}_z \tau$, while the second is given by $\hat{m}_q = S_z \tau$, obtained from the curl relation $i\hat{\mathbf{k}}_{\perp} \times \mathbf{q} = 2\boldsymbol{\Omega}_{\tau}$.

Accordingly, the medium-induced Lagrangian simplifies to
\begin{equation} \label{eq:Lagrangian_Q}
    \mathcal{L}_Q =  \overline{\psi} \hat{S}_z \tau  \psi= \sum_{\lambda} \psi^{\dagger}_{\lambda} \lambda \tau \psi_{\lambda},
\end{equation}
which describes the interaction between light’s polarization and its extrinsic orbital degrees of freedom ~\cite{yang2025_inducedBerry}. This term, known as the spin–orbit coupling~\cite{Liberman1992SpinorbitIO,Bliokh2015SpinorbitIO}, originates from the first-order derivative properties of the medium itself.

In the following, we demonstrate that Eq.(\ref{eq:Lagrangian_Q}) and Eq.(\ref{eq:Lagrangian_A}) are equivalent representations in the case of helical optical fiber systems. Consider an initial polarization vector of the displacement field $\mathbf{D}$ aligned along the $x$-axis, expressed as $ {\bf e}_x = {\bf e}_+ +{\bf e}_-$. Due to the geometric phase effect, the components of the right and left circular polarization acquire phase factors of opposite sign, ${\bf e}_{\pm}\rightarrow e^{\mp i\gamma(s)} {\bf e}_{\pm}$. Accordingly, we introduce a new set of polarization vectors by a rotation, $({\bf e}^{\prime}_x,{\bf e}^{\prime}_y)^T=R(\gamma(s))({\bf e}_x,{\bf e}_y)^T$, 
where $\gamma(s)$ denotes the Berry phase, corresponding to the rotation angle of the polarization plane ~\cite{Chiao1986ManifestationsOB,Vinitskii1990REVIEWSOT}. These rotated vectors, ${\bf e}^{\prime}_x$ and ${\bf e}^{\prime}_y$, can be identified with the transformed electric and magnetic field components, $\mathbf{D}^{\prime}$ and $\mathbf{B}^{\prime}$, respectively. As a result, the electric field, magnetic field, and propagation direction form a complete orthonormal tetrad, defining a three-dimensional local reference frame. Here, the propagation direction corresponds to that of the photon. Therefore, the dual transformation defined in Eq.~(\ref{DU}) is formally analogous to a rotation, where the rotation angle $\theta(s)$ is naturally identified with the geometric phase $\gamma(s)$. This relationship can be expressed as ~\cite{Kugler1987}
\begin{equation} \theta(s) =\gamma(s)= \int \Omega(s) ds. \end{equation}
where $\Omega(s)$ represents the angular velocity along the propagation path. Notably, when light propagates along a helical trajectory, the angular velocity corresponds to the torsion $\tau$ of the path, $\Omega(s) = \tau$ for a uniformly helical waveguide. 
The redefined field vectors $\mathbf{D}^{\prime}$ and $\mathbf{B}^{\prime}$ undergo parallel transport, which is equivalent to a transformation from a non-inertial reference frame to an inertial one ~\cite{daCosta1981QuantumMO,Kugler1987,Bliokh2009}.The gauge field serves as the connection governing parallel transport in this formulation, which leads to the vanishing of the effective gauge field in the inertial frame:
\begin{equation}\label{eq:gauge-torsion}
{\mathcal{A}}_s + \partial_s \gamma(s) = 0 \quad \Rightarrow \quad {\mathcal{A}}_s = -\tau. 
\end{equation}
It follows directly from \eqref{eq:Lagrangian_Q},\eqref{eq:Lagrangian_A} and \eqref{eq:gauge-torsion} that the medium-induced Lagrangian $\mathcal{L}_Q$ and the gauge field Lagrangian $\mathcal{L}_{\mathcal{A}}$ are identical, $\mathcal{L}_{\mathcal{A}} = \mathcal{L}_Q$. Noting that the torsion $\tau$ arises from the dipole moment, it can be interpreted as the extrinsic orbital angular momentum (EOAM). This equivalence demonstrates that spin–orbit coupling emerges naturally as a consequence of local dual symmetry, becoming particularly evident in geometrically structured systems such as helical optical fibers.

Since photons can also carry intrinsic orbital angular momentum (IOAM), $\hat{\bf L}=\hat{\bf x}\times\hat{\bf k}$, we generalize the photon wave function to a twisted optical field, written as  $\psi^+ e^{i l \varphi}$ and $\psi^- e^{-i l \varphi}$, where $l$ is the topological charge and $\varphi$ is the azimuthal angle ~\cite{Allen1992OrbitalAM}.For such a twisted beam, an additional first-order correction arises from $\mathbf{q}\cdot  \hat{\mathbf{k} }_{\perp} =\hat{L}_z \tau$ in $\Delta\hat{m}_0$, which yields
\begin{equation}
        \mathcal{L}_L =  \overline{\psi} \hat{L}_z \tau  \psi= \sum_{\sigma} \psi^{\dagger}_{\lambda_\sigma} l \tau \psi_{\lambda_\sigma},
\end{equation}
where $\psi_{l+1} \equiv (\psi_1e^{il\varphi},0,0,\psi_4e^{-il\varphi})^T$ and $\psi_{l-1}=(0,\psi_2e^{il\varphi}, \psi_3e^{-il\varphi},0)^T$ represent the helicity eigenstates modified by a vortex phase. We denote these states collectively as $\psi_{\lambda_\sigma}$.
Here, the total helicity $\lambda_{\sigma}$ is defined as
\begin{equation} 
\lambda_{\sigma} = \mathbf{J} \cdot \frac{\mathbf{k}}{k}, 
\end{equation}
which, for a vortex beam with topological charge $l$, becomes $\lambda_{\sigma}= (l + \sigma)$.
This expression shows that, although the interactions of polarized light carrying topological charge $l$ with the medium appear formally on an equal footing~\cite{Bliokh2006}, their underlying physical mechanisms are fundamentally different. In particular, the contribution associated with spin-orbit coupling originates from duality symmetry.

\section{Conclusion}
We investigate the localization of the dual symmetry of the electromagnetic field in helical systems using a field-theoretical framework, in which a novel gauge field emerges naturally as a consequence of the local symmetry. We find that this dual gauge field renders the coordinates of the photon non-commutative, thereby revealing the fundamental nature of the gauge field responsible for the localization of photons: it serves as the bridge connecting the canonical coordinate and the physical coordinate. The spin–orbit interaction arising during beam propagation along helical trajectories is formally equivalent to the coupling between the helicity and the dual gauge field. This correspondence provides a unified gauge-theoretic interpretation of the spin Hall effect of light, where such phenomena can be viewed as analogues of axion–electromagnetic coupling~\cite{Wilczek1987TwoAO, Gasperini1987AxionPB, Nikitin2012SymmetriesOF, Visinelli2013AxionElectromagneticW}. For vortex beams, corrections from orbit–orbit interactions remain significant, yet these are fundamentally distinct from spin–orbit coupling, which originates from a deeper intrinsic symmetry. This distinction provides a fresh perspective for differentiating between the spin and intrinsic orbital angular momentum (IOAM) of light.

A particularly important result is Eq.~(\ref{eq:Lagrangian_A}). In conventional electrodynamics, it is well known that charge conservation arises naturally from the $U(1)$ gauge symmetry of the electronic system, and that promoting this symmetry to a local one introduces the electromagnetic gauge potential—enabling interactions between electrons via photon exchange. In a completely analogous manner, the dual symmetry considered here is also a $U(1)$ symmetry, whose associated conserved quantity is helicity. Localizing this dual symmetry introduces a \textbf{dual gauge field}, implying that photons interact with each other through the exchange of quanta associated with this field. Consequently, our framework may provide a fundamentally new theoretical basis for describing interactions between light and matter.

Although the present work focuses primarily on reinterpreting the spin–orbit interaction, it is also valuable to explore the connection between non-Abelian gauge fields and the medium when this dual symmetry is extended in an isospin-like manner. Furthermore, it is interesting to consider the physical consequences of breaking this dual symmetry, such as the lifting of spin degeneracy and the emergence of a $\gamma^5$ term.

The work was supported by the National Key R$\&$D Program of China through Grant No. 2020YFC2201400, and the National Natural Science Foundation of China through Grant No. 12375084.

\nocite{*}

\bibliography{Duality}

\end{document}